# Cognition Engines: A Row-Scale HVDC Architecture for Computational Continuity of AI

## Version & Control

- Version: **v0.0**
- Date: **September 15, 2025**
- Author: **Paul Churnock, 4th Dimension Innovation (Independent Research)**

## Abstract

AI training and inference workloads create synchronized, step-dominant demand spikes that destabilize datacenter power systems and threaten cognition at millisecond timescales. Choukse et al. [2025] measured these dynamics directly, showing that large-scale training jobs spanning tens of thousands of GPUs produce tens of megawatts of oscillations in the 0.2–3 Hz band. This paper is an architectural positioning, not a deployment report: we propose a row-scale ±400$V_{dc}$ Cognition Engine that integrates Dynamic Response Units (DRU), solid-state transformers (SST), and a thin fast-reserve plane to preserve *Computational Continuity* under such events. DRUs inject fast energy via controlled droop and kHz-class inner loops; SSTs regulate average power under bounded ramps while enforcing no reverse power flow and no high-frequency export at the PCC; import is subject to a bounded dP/dt envelope. Distributed film capacitance and clamps absorb the first edge.

The row contract is measurable: ±1% steady band, ≤ 2% transient deviation, ≤ 50ms recovery, ≥ 45° equivalent phase margin, and reserve floors maintained in both directions, verified at the spine and two worst branches. Recharge is strictly valley-following: DRUs are restored only when instantaneous demand lies below a flat $P_{avg}$, MV headroom exists, and reserve floors are safe; commands are ramp-bounded (≈ 50 kW/s per row) so neither the DC rails nor the feeder sees the training waveform. Protection is time-graded: branch interruption in microseconds, row sectionalization in milliseconds, MV FLISR in seconds.

Scaling preserves the invariant. Pods add slow schedulers to allocate averages, halls add feeder amp-cap enforcement and FLISR, campuses add transmission arithmetic; at no stage is the row contract diluted or retuned. Compliance is demonstrated by waveform evidence, with branch faults clearing in microseconds, rows holding inside the 2% / 50ms window, MV loops restoring with no reverse power flow and no high-frequency export at the PCC, and synchronized bursts absorbed without ripple. With these traces, continuity moves from aspiration to specification.

The architecture maintains continuity under single-element loss (N+1 redundancy) within a defined fault envelope, establishing a structural alternative to legacy UPS/STS approaches. Anchored in measured AI workload dynamics, it defines the architectural blueprint for sustaining Computational Continuity under the scaling regimes of frontier training and inference systems.

## Thesis Claims & Contributions

This paper advances a new doctrine for AI-scale power: continuity is preserved not by smoothing loads or extending legacy UPS/STS, but by re-architecting distribution itself so that cognition persistence becomes a structural property. Grounded in empirical workload traces of synchronized, MW-scale oscillations in the 0.2-3Hz band [Choukse et al., 2025], this paper extends beyond incremental mitigations and defines a new substrate for AI-native infrastructure.

This paper makes five contributions:

1. **Workload-Coupled Instability Modeling.** We model the instability of AI/LLM training workloads at the row-bus level, capturing their step-dominant, synchronous power swings, and analyze the limitations of legacy UPS/STS systems under correlated surges.

2. **Solid-State Distribution Architecture.** We propose a row-scale ±400$V_{dc}$ system that couples Dynamic Response Units (DRUs), solid-state transformers (SSTs), DC-layer protection, and MV-FLISR to form an adaptive, self-healing electrical mesh.

3. **Workload-Aware Recharge Logic.** We redefine valley-following recharge to incorporate multi-scale temporal averaging, blackout of communication phases, and state-of-charge urgency scaling, preventing recharge-induced instability while maintaining average power alignment.

4. **Architectural Stability Through SST Wave-Shaping.** We demonstrate that bi-directional SSTs, beyond serving as row gateways, can actively shape impedance and present deterministic "golden row" signatures to the MV feeder, transforming multi-row scaling from a coordination burden into electrically independent stability.

5. **Protection and Control with Lifecycle Margins.** We reframe protection from idealized microsecond selectivity to engineered coordination with explicit lifecycle margins, discrimination between structured training bursts and chaotic fault signatures, and defined degradation modes in the Fast-Reserve-Plane (FRP).

Collectively, these contributions demonstrate analytically that the proposed architecture expands dynamic stability margins compared to current practice, while defining an explicit framework for scaling, deployment, and eventual economics evaluation.

# 1. Introduction: From "Nines" to Computational Continuity

## 1.1 Problem Statement & Infrastructure Gap

AI training behaves like a synchronized machine. Recent analysis of large-scale training workloads spanning tens of thousands of GPUs [Choukse et al., 2025] documents step-dominant power excursions with edges in tens of milliseconds tied to batch cadence, where compute-heavy phases approach GPU thermal limits while communication-heavy phases collapse toward idle draw. These oscillations are not minor, manifesting at scale as tens to hundreds of megawatts swinging in the 0.2-3Hz band, directly overlapping with turbine-generator and grid resonance modes. This paper is written as a solution response: where *Choukse et al.* established the synchronized, step-dominant workload envelope, we present an implementable row-scale power architecture that preserves Computational Continuity under those same conditions.

The mechanism is electrical: GPU front ends and VRMs have finite holdup and bandwidth, and constant-power behavior converts a bus dip into a surge that deepens the dip. When sag magnitude or duration exceeds margin, the result is resets, link flaps, throttling, or job preemption. With racks at hundreds of kilowatts and rows at megawatt scale, even modest correlation across hundreds of racks produces step swings at tens to low hundreds of megawatts within tens of milliseconds. It is precisely this correlation (synchronous by design AI training) that invalidates assumptions of statistical diversity embedded in legacy datacenter design.

We define acceptance in these terms: stability must hold within ≤2% bus deviation and ≤50ms recovery to the steady band, measured at the row spine and the two electrically worst branches. These thresholds are drawn from power systems engineering practice for constant-power loads and the documented failure modes observed in large-scale AI training deployments.

## The Infrastructure Gap

Industry approaches focus on improving rack conversion efficiency and density, but they do not coordinate aggregate steps across shared infrastructure. Medium-voltage distribution plants were designed for slower and statistically diverse loads, not for synchronized constant-power steps. The gap is visible at three interfaces:

- Medium-voltage distribution must ride through coordinated steps without breaching segment or power-quality limits;

- The utility interconnect must absorb facility swings without export or disturbance to neighboring systems; and
- Protection must discriminate load steps from faults, requiring selective DC interruption at the branch, sectionalization at the row, and medium-voltage restoration that is time-graded and non-conflicting.

Per-rack storage and PSUs cannot enforce these constraints at scale. They act without visibility to segment limits, they amplify constant-power feedback, and they cannot pool energy across the row. Utility-scale batteries and grid-forming inverters operate effectively on minute-to-hour timescales, but not at the millisecond bursts that dominate row dynamics.

This is the structural gap our architecture addresses. The row DC bus must be treated as the atomic grid: a domain where DRUs enforce stability through workload-aware recharge, SSTs shape row signatures to decouple multi-row dynamics, protection layers operate with lifecycle margins, and the FRP provides continuity even under network partitions. By focusing on the millisecond physics of correlated steps, Computation Continuity becomes a designed property rather than a probabilistic outcome.

Facility requirements are concrete: medium-voltage segments must respect current caps, maintain no reverse power flow and no high-frequency export at the PCC, and comply with flicker, harmonic, and fault-duty limits while keeping the row bus inside the acceptance band during burst trains. In this framing, electrical acceptance is not about uptime alone but about preserving Computational Continuity under AI-scale demand.

## 1.2 Computational Continuity Framework & Why the "Nines" Fail at Millisecond Scale

Calendar uptime does not protect millisecond events. What matters is whether computation continues through electrical disturbance. We formalize this as *Computational Continuity*, defined as the ability of a datacenter row to sustain workloads without resets, link flaps, throttling, or job aborts when subjected to prescribed electrical disturbances.

Traditional availability metrics aggregate outages into hours or minutes. Computational Continuity fails in milliseconds. A facility can claim five nines of availability while bus transients still trigger resets, throttling, or multi-hour job aborts. This is not hypothetical; telemetry from large-scale AI training shows that synchronous jobs produce coordinated step excursions with edges on the order of tens of milliseconds, exactly where traditional uptime metrics are blind. Such statistics conceal the physics of step-dominant constant-power loads, converter bandwidth, and short-duration reserves. At megawatt densities with ±400V racks, the governing variables are energy reserve and voltage regulation, not calendar uptime. The relevant failure mode is no longer an isolated server reboot. It is a synchronized disruption across thousands of GPUs, producing a collapse in active training or inference.

Computational Continuity is satisfied only if fast reserve margins remain above minimum thresholds during the test window. We denote these reserves as $R\uparrow$ (upward response capacity, measured in kilowatts available for injection) and $R\downarrow$ (downward absorption capacity, measured in kilowatts of headroom available for sink).

The acceptance criteria are explicit. The row DC bus must remain within two percent of nominal during any step profile in the design envelope and must return to the steady band within fifty milliseconds, with $R\uparrow$ and $R\downarrow$ above their floors. Measurements are taken at the row spine and at the two electrically worst branches determined after installation by impedance. Conformance is proven through disturbance waveforms that mirror the step-dominant traces documented in synchronized training workloads, ensuring discrimination between structured bursts and actual faults.

This reframes reliability from an abstract uptime metric into a testable property of the electrical system, making the row the control surface where fast elements handle millisecond physics and average power is scheduled against medium-voltage limits. In this framework, "nines" are accounting artifacts, and Computational Continuity is engineering reality.

## 1.3 System Architecture & Boundary

The target is a row that can ride through synchronized bursts within the design envelope while meeting both Computational Continuity and medium-voltage coordination requirements. Protection must remain selective inside this envelope, with branch-level clearing and row sectionalization as specified. Outside the envelope, the system must degrade deterministically: affected segments isolate cleanly, uninvolved segments continue to operate, and the failure trajectory is bounded rather than cascading.

Success is defined in terms of Computational Continuity, not calendar availability. A row meets the standard only if prescribed disturbance waveforms can be applied without triggering resets, throttling, or job aborts, and if reserves remain above floor throughout the test window. Conformance is demonstrated through standardized waveform acceptance and workload verification, supplemented by periodic field audits. This reframes reliability from an abstract uptime metric into a testable property of the electrical system.

This standard reframes the system boundary as the locus of millisecond physics; where DRUs, SSTs, and protection act in concert to absorb AI workload oscillations without breaching either row stability or MV compliance.

**System Boundary**

The row is the atomic grid boundary. Inside this boundary are the $\pm 400 V_{dc}$ floating spine, Dynamic Response Units (DRUs) that provide seconds-scale reserves, bi-directional solid-state transformers (SSTs) that regulate average power, and reserve coordination that is Tier 0 by configuration with a supervised tier governing valley-following recharge scheduling.

Protection within the row includes solid-state branch interruption, hybrid sectionalization, and insulation monitoring with supervised reclose. Protection is specified with lifecycle margins, ensuring coordination holds under device tolerance, thermal drift, and long-term aging rather than idealized microsecond intervals.

The medium-voltage interface is defined by voltage class, power rating, segment current limits, no reverse power flow and no high-frequency export at the PCC, compliance with utility flicker and harmonic limits, and bounded fault-current contribution with telemetry. Outside this scope are standby systems, long-duration storage, campus interties, and building support systems. They connect to the row only through defined interfaces for voltage, power, control, protection, electromagnetic compatibility, isolation, and power quality.

This boundary ensures that acceptance testing, and continuity guarantees are scoped where failures occur: in the millisecond dynamics of the row. It also defines the FRP degradation path; Tier 0 analog continuity, Tier 1 partial clusters, and Tier 2 full coordination, so that failure modes remain explicit and bounded.

Above this boundary, traditional availability metrics remain applicable at the facility-distribution layer, where traditional multi-source utility and generator backup govern minute-to-hour resilience.

**Workload Model**

Training workloads are step dominant. For design and test, we model bursts at 10 to 25 percent above average load, lasting 10 to 90 seconds, with edge transitions of 100 to 800 milliseconds. Peak-to-average ratios range from 1.1 to 1.25. Rack timing correlation (ρ) is modeled between 0.2 and 0.6 across a job. Idle power raises the baseline but does not eliminate steps. Orchestration aligns batch boundaries and shapes edge timing only; it does not remove aggregate step behavior.

The row must absorb these bursts locally, so Computational Continuity is preserved, and medium-voltage limits are not exceeded. DRU energy and control are sized from this envelope, with

$$E_{\text{dru}} \geq P_{\text{row}} \times T_{\text{bridge}}$$

and recovery required within 50 milliseconds. This workload is consistent with the power oscillation characteristics documented in large-scale AI training deployments. Recharge follows the valley-forming discipline defined in §3.5, ensuring that DRU restoration never disturbs the DC bus or MV feeder.

This workload envelope is consistent with the power oscillation characteristics documented in large-scale AI training deployments [Choukse et al., 2025], ensuring that design and acceptance reflect observed training behavior rather than assumed averages.

**1.4 Architectural Answer (Preview)**

A ±400V$_{dc}$ floating row bus reduces current and copper while defining the atomic grid. Dynamic Response Units (DRUs) supply fast energy through controlled source impedance with droop and bounded slew. Inner-loop bandwidth is in the kilohertz range with defined phase margin, and DRUs are sized for a seconds-scale bridge time ($T_{bridge}$). Solid-state transformers (SSTs) regulate average power with bounded ramps, bidirectional flow, and a small positive droop to support damping and load sharing.

Protection is time-graded: branch faults clear in microseconds, row sectionalization occurs in milliseconds, and medium-voltage restoration follows in seconds. SSTs provide more than conversion; they shape impedance and enforce deterministic "golden row" signature at the MV interface, decoupling row dynamics from facility resonance. The Fast-Reserve Plane (FRP) defines explicit degradation modes so that even under network partitions or controller loss, Tier 0 continuity is guaranteed, and recovery is bounded. DRU recharge follows the valley-forming discipline defined in §3.5, and lifecycle margins are specified once in §2.1.

This architecture enables the row to absorb synchronized bursts locally while sustaining Computational Continuity and meeting medium-voltage coordination requirements. Continuity is treated as an invariant of the design, verified through acceptance testing rather than availability statistics. This preview frames the design approach developed in the following sections, where each element is expanded into specification, verification, and validation detail.

### 1.5 Related Work

Choukse et al. [2025] provide the clearest telemetry to date on the electrical footprint of large-scale AI training. They show that synchronized workloads spanning tens of thousands of GPUs produce megawatt-scale oscillations concentrated in the 0.2–3Hz band, overlapping turbine-generator and grid resonance modes. Their diagrams make the case unambiguous: the workload itself is the destabilizing agent. This paper builds directly on those measurements, positioning an architectural response to sustain Computational Continuity under the exact dynamics they documented.

Classic studies of power system stability emphasize electromechanical oscillations, inter-area modes, and turbine-governor dynamics [Kundur, 1994; Pai, 1989], with little attention to millisecond disturbances originating from compute. Distribution-level work on DC microgrids [Lasseter, 2002; Schonberger et al., 2006] has focused on renewable intermittency and droop-based sharing, but not on correlated step-dominant bursts. The scaling regime of AI clusters presents a fundamentally different forcing function—periodic, synchronized, and coupled across thousands of devices.

Research on solid-state transformers (SSTs) [She et al., 2013; Wang et al., 2020] highlights their ability to regulate power quality, provide bidirectional conversion, and embed fast protection. These results reinforce the SST role adopted here: the single galvanic isolation boundary and controllable gateway between medium-voltage and row-scale DC.

Finally, datacenter reliability literature has long been defined by availability "nines" [Barroso et al., 2017], typically achieved through UPS/STS replication. These mechanisms preserve uptime in the face of generator or utility loss but do not address millisecond-scale destabilization. Our work reframes reliability in terms of Computational Continuity: waveform compliance at the row boundary, with availability metrics remaining relevant only at the campus-distribution layer where multi-source utility and generator redundancy apply.

This paper is the first to define Computational Continuity as an architectural invariant, directly addressing the workload dynamics quantified by Choukse et al.

## 2. Architectural Overview

The row serves as the atomic control boundary, with three integrated components: physical infrastructure, functional control elements, and defined operational states.

### 2.1 Physical Stack

Medium-voltage AC feeds solid-state transformers (SSTs), which convert utility power to a regulated ±400V$_{dc}$ floating spine to each row. Racks connect by short HVDC branches to rack power distribution units (PDUs) that complete conversion to 48V$_{dc}$. Galvanic isolation is enforced at the SST interface, so downstream DC-DC stages are not required to provide additional isolation except where dictated by regulatory or topology constraints.

Dynamic Response Unit (DRU) cabinets couple directly to the spine, contributing fast energy through droop sharing with kilohertz-class inner-loop bandwidth. Seconds-scale reserve is sized by the bridge time $T_{bridge}$, with energy capacity $E_{tot}$ (detailed in Section 3). The spine itself is laminated copper with touch-safe bus connections, mandatory pre-charge on every close, and continuous monitoring of insulation resistance using an Insulation Monitoring Device (IMD).

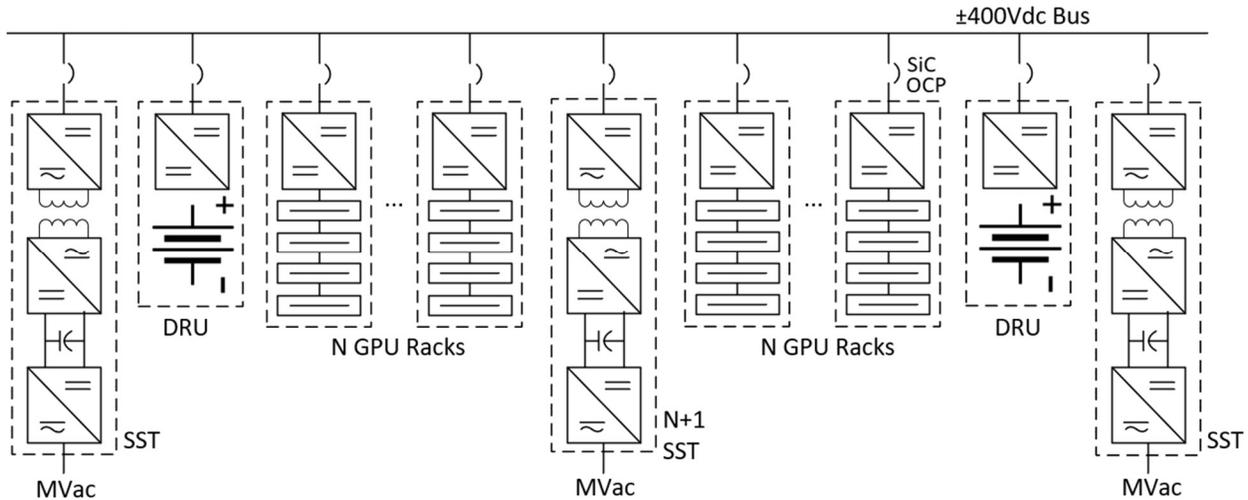

Figure 1 - Row-scale ±400V$_{dc}$ topology

The stability architecture emerges from the component-level control characteristics rather than software coordination. Solid-state transformers operate as bidirectional voltage-controlled current sources (VCCS) with inherent positive droop, naturally providing damping and average power regulation. Dynamic Response Units function as current-controlled voltage sources (CCVS) with fast current response to voltage deviations, creating immediate bus stiffening. This VCCS/CCVS pairing establishes the control hierarchy through analog feedback loops: DRUs respond to voltage errors in microseconds through their current control characteristics, while SSTs respond to power setpoints in seconds through their voltage control with droop. The result is a physically stable system where fast dynamics are handled by current sources responding to voltage, and slow dynamics are managed by voltage sources responding to power commands.

Beyond average power regulation, SSTs actively shape row impedance and enforce deterministic "golden row" signatures at the MV feeder. This decouples row-level SoC dynamics from facility-scale resonance, enabling independent row operation without complex synchronization protocols.

This branch-and-spine topology minimizes distribution losses while enforcing defined isolation boundaries.

Protection is applied at the points where faults originate. Each branch is equipped with solid-state interruption, the spine incorporates sectionalizers that use solid-state commutation with mechanical isolation, and conventional relaying is maintained at the medium-voltage interface. Protection settings are defined with lifecycle coordination margins, ensuring selectivity under device tolerance spreads, thermal drift, and end-of-life degradation.

Thermal integration is explicit. DRU cold plates tie into the row liquid loop, and the cooling budget is specified to accommodate cyclic energy exchange at seconds-scale reserve.

Recharge coordination is not embedded at the physical layer but governed by the reserve framework in §3.5. FRP hooks at the SST–DRU interface ensure that even if higher-level coordination is lost, Tier-0 analog control preserves continuity.

This physical stack defines the hardware envelope within which functional roles are realized.

## 2.2 Functional Roles

The system achieves stability through a control hierarchy operating at different time scales:

**Dynamic Response Units (DRUs).** Operating as current-controlled voltage sources (CCVS) supply fast energy and stiffen the bus through droop control ($r_{dru}$) and a bounded slew rate $|dP/dt|_{dru}$, correcting millisecond-scale errors. They hold state of charge within a mid-band $[L_1, L_2]$ and do not chase average energy; the bridge time $T_{bridge}$ sets seconds-scale reserve. Parallel DRUs reduce droop and stiffen the bus proportionally to count. Recharge discipline follows the valley-following framework detailed in §3.5, ensuring reserves are restored without destabilizing the bus. These slew-rate ceilings directly constrain energy cell specification, linking architecture to chemistry.

**Solid-state Transformers (SSTs).** SSTs are bidirectional, grid-following voltage-controlled current sources (VCCS). They regulate average power with bounded ramp rate $|dP/dt|_{sst}$ and apply a small positive droop ($r_{sst} \ll r_{dru}$) to support damping and sharing. SSTs enforce no reverse power flow and no high-frequency export at the PCC; import is subject to a bounded dP/dt envelope and maintain medium-voltage segment caps. In addition to regulation, SSTs actively shape row impedance and present deterministic "golden row" signatures at the MV feeder, decoupling row dynamics from facility-scale resonance.

**Rack PDUs.** Rack PDUs serve as limited-slew, power-capped front ends, defined by ($S_{pdu}, P_{pdu}$), and are verified in acceptance testing.

**Protection.** Protection follows fault origin. Solid-state branch interrupters bound fault energy to $E_{clamp,max}$ and clear within $t_{clear,branch}$ at the worst-case loop inductance $L$. Spine sectionalizers establish a current zero and isolate within $t_{iso,row}$. Medium-voltage relays provide restoration on the seconds scale. Settings incorporate lifecycle coordination margins defined in §2.1, ensuring selectivity under semiconductor tolerance, sensor drift, and device aging. This time-graded approach ensures selective operation: the fastest devices clear first, with slower upstream devices providing backup (detailed in §7).

**Stability Arises in Layers.** Tier-0 stability is enforced by droop and current limits at the bus; it remains stable even if signaling fails. Tier-1 stability is time-synchronized and cryptographically signed, coordinating reserve scheduling, valley-following recharge, and inter-row assists under segment caps. The Fast-Reserve Plane (FRP) defines explicit degradation paths; Tier 0 analog continuity, Tier 1 cluster coordination, Tier 2 full networked response, so failures degrade deterministically rather than cascade.

**Control Layering:** The DRU inner loop (kHz) regulates row current/bus stiffness; a seconds-scale energy manager enforces amplitude and dP/dt; the PCC controller (hundreds of ms to seconds) enforces bounded import and no reverse power flow. Objectives are orthogonal, preventing loop contention.

These functional roles map the physics of each component into the layered stability model.

## 2.3 Power-Flow States

Row behavior is defined in operating states.

**Normal Operation.** The spine remains within the Computational Continuity (CC) steady band (±1%). DRUs operate in the mid state-of-charge range $[L_1, L_2]$ and carry noise-floor variations $\Delta P$ under droop $r_{dru}$ with a slew ceiling $|dP/dt|_{dru}$. SSTs regulate average power with ramp limit $|dP/dt|_{sst}$ and a small positive droop $r_{sst}$, while also enforcing impedance shaping at the MV interface, presenting deterministic "golden row" signatures that decouple row dynamics from facility resonance.

**Synchronized Step Response.** When training workloads create synchronized power steps (documented in [*Choukse et al., 2025*]), distributed bus capacitance at branches and DRU terminals supplies the first 50–150 μs. This behavior corresponds directly to the burst envelope characterized in Choukse et al.; the sequence below shows how the row-scale topology provides the structural solution response, sustaining Computational Continuity within the documented thresholds. DRUs then take the remainder with kilohertz-class inner-loop bandwidth and ≥45° margin, presenting controlled source impedance that damps the spine pole formed by bus capacitance ($C_{bus}$) and loop inductance ($L$). GPU input rails remain within PSU regulation and holdup. Bus deviation is limited to ≤2% and settles within 50 ms, satisfying CC. If Tier 1 coordination is lost, Tier 0 analog physics alone sustains this stability, reflecting FRP degradation guarantees.

**Recharge Coordination.** A Tier-1 scheduler admits recharge only when MV headroom and reserve floors are intact. Admission is workload-aware, filtered to exclude communication phases, and urgency-scaled by SoC, but the detailed framework is defined in §3.5. If coordination is lost, Tier-0 physics sustains stability.

**Medium-Voltage Contingency.** FLISR isolates the fault in ~1–3 s; DRUs bridge for the duration of $T_{\text{bridge}}$. SSTs rebalance through a common assist setpoint with droop sharing and rate limits. Each row's assist is capped at $P_{\text{assist,max}}$, preserving segment limits and ensuring no reverse power flow and no high-frequency export at the PCC.

**Fault Conditions.** Branch shorts clear within $t_{\text{clear,branch}}$ with clamp energy $\leq E_{\text{clamp,max}}$. A sectionalizer isolates only if a segment fault persists. Insulation Monitoring Devices (IMDs) localize faults to branch or segment level, and the system trips the smallest island required to restore insulation margin, followed by supervised reclose. Uninvolved segments remain in service. Protection margins are defined by lifecycle coordination in §2.1, ensuring selectivity under component tolerances, drift, and aging.

These power-flow states formalize how the row behaves under operating, recharge, and fault conditions, setting up the architectural synthesis that follows. The row is now defined in stack, roles, and states. Stability is structural, enforced by physics and verified through testing. Section 3 develops the models that prove it.

## 3. First-Principles Models (Math You Can Design To)

We model the row as a regulated ±400$V_{\text{dc}}$ spine (|V|≈400V per rail, 800V differential) fed by solid-state transformers and stiffened by a bank of Dynamic Response Units. The variables used repeatedly are:

| Symbol | Meaning | Units |
|---|---|---|
| $P_{\text{row}}$ | Row Nameplate Power | kW or MW |
| $P_{\text{avg}}$ | Average draw of Row | kW or MW |
| $\Delta P(t)$ | Synchronized overage (burst above average) | kW or MW |
| $v(t)$ | Bus deviation from steady band | % or V |
| $E_{\text{tot}}$ | Total DRU stored Energy | kJ |
| $P_{\text{pk}}, P_{\text{cont}}, E_{\text{use}}$ | DRU per-shelf peak power, continuous power, usable energy | kW, kW, kJ |
| $SoC \in [0,1]$ | DRU State of Charge | Fraction |
| $\omega_{\text{dru}}, \omega_{\text{sst}}$ | DRU inner-loop and SST control loop bandwidth | rad/s |

This section supplies the first-principles inequalities that make **Computational Continuity** a testable property, directly resolving the workload instability envelope documented in *Choukse et al., 2025*.

### 3.1 Load Model for Training Bursts

Training behaves like a synchronized step added to a steady base, requiring control approaches specifically designed for AI workload characteristics. Row power is written as

$$P_{\text{row}}(t) = P_{\text{avg}} + \Delta P(t), \qquad \Delta P(t) = \alpha(t)\, P_{\text{avg}},$$

with $\alpha(t)$ a bounded, piecewise-constant process capturing orchestration.

Energy to cover a surge is $E = \alpha P_{\text{row}}$. Example: $\alpha = 0.25$, $P_{\text{row}} = 1$ MW, $\tau = 60/90$ s $\Rightarrow E = 4.17/6.25$ kWh. DRU reserves of tens of kWh per row therefore close seconds-scale envelopes without bulk storage.

The worst-case envelope is a step of amplitude $\alpha_{max} \in [0.1, 0.25]$ sustained for a window $T_{\text{surge}} \in [10, 90]$ s, plus sub-second edges of rise time $\Delta t_{\text{edge}}$ shaped by the GPUs. The energy behind one surge is

$$\Delta E = \int_0^{T_{\text{surge}}} \Delta P(t)\, dt \approx \alpha_{max}\, P_{\text{avg}}\, T_{\text{surge}}.$$

This envelope corresponds directly to the instability profile documented in *Choukse et al., 2025*. For DRU sizing, the bridge condition is enforced as:

$$E_{\text{tot}} \geq \Delta E \quad \text{Power Gate:} \quad N\,P_{pk} \geq \Delta P \text{ ; } \textit{Slew gate:} \quad |dP/dt|_{dru} \geq \Delta P/\Delta t_{edge}$$

Ensuring both the energy and slew capability exist to absorb synchronized steps without bus collapse. The slew-rate ceiling $|dP/dt|_{dru}$ is explicitly bounded and constrains DRU cell specification, linking the architecture to chemistry.

Recharge behavior is governed by the valley-following framework detailed in §3.5, ensuring that recovery occurs only during validated valleys and never destabilizes the bus. SSTs enforce medium-voltage compliance not only through ramp limits but also by presenting controlled "golden row" impedance profiles, decoupling row-level recharge dynamics from facility-scale resonance.

Everything that follows ensures the bus deviation stays bounded while $\Delta P(t)$ is absorbed by the DRU bank rather than pushed to the medium-voltage feeder, establishing Computational Continuity as a structural property of the row. SSTs maintain MV compliance not only through ramp limits but also by presenting controlled "golden row" impedance profiles that decouple row-level dynamics from facility-scale resonance.

### 3.2 DRU as Fast Bus-Former

We do not claim spectral cancellation of the 0.2–3 Hz fundamental; the row enforces bounded presentation at the PCC (tight bus limits and explicit bounds on |ΔP| and dP/dt) so rack-level pacing and utility controls operate without local instability.

Each shelf is a bidirectional power stage with tight current control and finite energy. As baseline per shelf we take:

$$P_{pk} = 40 \text{ kW (40 s)}, \quad P_{cont} = 24 \text{ kW (90 s)}, \quad E_{use} \approx 0.6 \text{ kWh}$$

With $N$ shelves, the hard constraints are:

$$\text{Power gate:} \quad \Delta P_{max} \leq N\,P_{pk}, \quad \text{Energy gate:} \quad \Delta E \leq N\,E_{use}, \quad \text{Thermal gate:} \quad \langle P_{dru} \rangle_T \leq N\,P_{cont}.$$

State of charge evolves as $\dot{\text{SoC}} = -(P_{dru}/E_{tot})$ with $E_{tot} = N\,E_{use}$. We operate in a mid-band SoC ∈ [SoC$_{min}$,SoC$_{max}$] = [0.5,0.8] so both source and sink headroom exist:

$$E_\uparrow = (\text{SoC} - \text{SoC}_{min})E_{tot}, \quad E_\downarrow = (\text{SoC}_{max} - \text{SoC})E_{tot}$$

Seconds-scale "fast reserve" floors follow:

$$R_\uparrow = \min(NP_{pk},\, E_\uparrow/T_\star), \quad R_\downarrow = \min(NP_{pk},\, E_\downarrow/T_\star),$$

for a design window $T_\star$ (chosen to match the longest credible burst or MV reconfigure time).

Loop behavior matters as much as nameplate. The DRU's inner current loop tracks bus-voltage error with bandwidth $\omega_{dru}$ in the kHz range, enforcing a static droop $r_{dru}$ [V/A]. With $N$ shelves in parallel, the aggregate droop $r_{eq} = \left(\sum_i r_{dru,i}^{-1}\right)^{-1}$, and the effective electrical stiffness $K_{bus} = 1/r_{eq}$ rises with population. This stiffness, together with the bus capacitance analyzed in §3.3, sets the small-signal pole that governs recovery after the synchronized steps documented in *Choukse et al., 2025*. The slew-rate ceiling $|dP/dt|_{dru}$ is an explicit design gate that constrains DRU cell specification, linking architecture directly to chemistry.

Recharge admission and scheduling are defined in §3.5; DRUs respond only to validated valleys, ensuring reserves are restored without destabilizing the same burst envelope they are built to protect.

### 3.3 Bus Dynamics, Film Capacitance, and Clamps

At the instant a step lands, the only actor is distributed capacitance $C_{bus}$ until the DRU loop engages. To limit the immediate dip to $\Delta V$ over a control latency $\Delta t$, the requirement is

$$C_{bus} \gtrsim \frac{I_{step}\,\Delta t}{\Delta V} \quad \text{with} \quad I_{step} \approx \frac{\Delta P_{max}}{V_{bus}}.$$

At 800 V differential, a 360kW row-step implies $I_{step} \approx 450$A. With $\Delta t = 75\mu s$ and $\Delta V = 0.02 \times 800 = 16$V, the target is $C_{bus} \approx 2.1$mF.

Film capacitance is sized to bound $\Delta V$ over the sub-millisecond first edge; DRUs handle ms–s energy, avoiding over-sizing caps that would introduce low-frequency resonances. This capacitance is distributed at branches and within DRU sidecars to minimize ESL/ESR.

Bus inductance $L_{bus}$ from cables and laminations sets the overvoltage on fast turn-off:

$$V_{ov} \approx L_{bus}\,(di/dt).$$

Snubbers or active clamps must be sized against the worst credible $L\,di/dt$ from a branch fault cleared in microseconds. Practical design treats these not merely as voltage limiters but as energy sinks with margin. Clamp sizing incorporates lifecycle factors; contact resistance growth, cable aging, and device tolerance, ensuring protection margins hold over service life rather than only at commissioning.

Once droop control is engaged, the linearized bus around its operating point behaves like a first-order system with pole

$$\omega_p \approx \frac{K_{bus}}{C_{bus}},$$

so $r_{dru}$ and $C_{bus}$ are tuned to place $\omega_p$ such that recovery completes within ≤50 ms and bus deviation returns inside ±1 %.

The capacitance and clamp design is therefore not an optimization variable but a compliance requirement; Computational Continuity holds only if immediate energy buffering and clamp absorption guarantee bounded deviation during the first tens of microseconds before DRUs engage.

### 3.4 SST as Slow, Damped Energy Source

From the DC side, a pure constant-power source has negative incremental conductance and can excite oscillations; we therefore add a small, positive droop $r_{sst} > 0$ and filter their setpoint with a long time constant $\tau_{sst}$. A practical law is:

$$P_{sst}(s) = \frac{P_0}{1+s\tau_{sst}} + \underbrace{\frac{1}{r_{sst}}}_{\text{damping}} v(s),$$

with $\tau_{sst}$ in seconds and $r_{sst}$ one to two orders larger (softer) than the DRU droop so that hierarchy is preserved.

In the time domain we enforce $|dP_{sst}/dt| \leq \rho$ (ramp limit) so the medium-voltage feeder sees only a bounded ramp rather than the training waveform. The SST remains bidirectional, with no reverse power flow and no high-frequency export at the PCC; import is subject to a bounded dP/dt envelope. Internally we allow $P_{sst} < 0$ briefly to absorb DC overshoot after a clearing event. Bidirectional operation is internal to the converter; reverse power flow is clamped at the PCC boundary.

Beyond average regulation, SSTs actively shape row impedance and present deterministic "golden row" signatures at the MV feeder, decoupling internal recharge and DRU dynamics from facility-scale resonance. This ensures multi-row impedance rather than synchronization risk.

If higher-level coordination is lost, FRP degradation ensures that SSTs default to Tier 0 analog damping with conservative droop and ramp, preserving continuity until full coordination is restored.

### 3.5 Reserve Accounting and Valley-Following Recharge

The fast-reserve plane operates on two numbers per row: instantaneous reserve floor $(R_\uparrow, R_\downarrow)$ in kW, and available headroom $(E_\uparrow, E_\downarrow)$ in kWh, both derived from SoC and the gating constraints above. Minimum floors are enforced:

$$R_\uparrow \geq R_\uparrow^{min} \text{ and } R_\downarrow \geq R_\downarrow^{min}.$$

Recharge is deferred when floors are at risk and admitted only when instantaneous row demand lies below the flat average target $P_{\text{avg}}$. In practice, "flat average" is replaced by workload-aware filters across iteration, phase, and job windows, with explicit blackout of communication phases and urgency scaling as SoC falls below critical bands. This prevents false valleys from triggering recharge during synchronized dips.

In this regime, no SST is reallocated from redundancy; $N+1$ remains reserved for continuity. The recharge command is arithmetic,

$$\textbf{Recharge law: } P_{chg} = \max\{0,\ \min[H_{mv},\ P_{avg} - P_{load} - R_{safety}]\}$$
$$\textbf{Feasible reserves: } \{(E_\uparrow, E_\downarrow) \mid E_\uparrow \geq R_\uparrow^{min},\ E_\downarrow \geq R_\downarrow^{min},\ E_\uparrow + E_\downarrow \leq E_{tot}\}.$$

where $H_{mv}$ is feeder headroom and $R_{safety}$ is a fixed margin preserving burst response. The command is rate-limited, so the DC bus voltage does not move and the SST ramps remain within relay and tariff bounds. Practical bounds are $|dP_{\text{chg}}/dt| \leq 50$ kW/s per row, with stricter limits if MV policy demands it.

The DRUs sink $P_{chg}$ until $SoC$ reaches $SoC_{\text{max}}$. Entry and exit are hysteretic in SoC (L1 < L2) and any reserve-floor violation vetoes recharge instantly.

By construction, recharge never compromises continuity, and reserves are restored without exposing the medium-voltage feeder to burst dynamics.

### 3.6 Sizing Synthesis (the Minimal Inequalities)

Given the workload envelope $(\alpha_{max}, T_{\text{surge}}, P_{\text{avg}})$ select the DRU population $N$ such that

$$N \geq \max!\left(\frac{\alpha_{max} P_{\text{avg}}}{P_{\text{pk}}},\ \frac{\alpha_{max} P_{\text{avg}} T_{\text{surge}}}{E_{\text{use}}}\right),\quad \langle P_{\text{dru}} \rangle \leq N P_{\text{cont}}$$

Choose the distributed capacitance $C_{\text{bus}}$ so the immediate dip satisfies $\Delta V/V_{\text{bus}} \leq 2\%$ over the DRU latency. Tune the DRU droop $r_{\text{dru}}$ so the composite pole yields a return to within ±1% in ≤50 ms.

Select $\tau_{\text{sst}}$ and $r_{\text{sst}}$ so the SST contributes damping without interfering with the DRUs, and cap $|dP_{\text{sst}}/dt|$ so the medium-voltage feeder obeys its relay and tariff ramp constraints.

Reserve accounting enforces minimum floors $R_\uparrow^{min}, R_\downarrow^{min}$ workload-aware recharge admission (iteration/phase/job filtering, comms blackout, SoC urgency scaling). Lifecycle safety factors are embedded in all four gates; tolerance, thermal, drift, capacity fade, are embedded in each inequality so that the contract holds over service life. FRP degradation guarantees Tier-0 continuity if higher-level coordination fails.

If those four lines hold—power, energy, thermal, and dynamics—the row will satisfy the Computational Continuity metric under any burst inside the design envelope of *Choukse et al., 2025*, while presenting as flat and well-behaved to the utility.

## 4. Row-Bus Stability Contract

The row's stability contract is device-agnostic and measured at the ±400$V_{\text{dc}}$ rails at the spine. It is enforced by physics before software:

**Continuity Limits.** In steady state, the bus holds ±1 %. Under any training step in the design envelope *Choukse et al., 2025*, deviation ≤2 % with recovery ≤50 ms, without hunting or subharmonics. Readings at the spine and two worst branches govern.

**Hierarchy.** DRUs act as current-controlled voltage sources (CCVS) with kHz loops and ≥45° phase margin, injecting current directly against voltage error; SSTs act as voltage-controlled current sources (VCCS) with seconds-scale ramp limits and positive droop, enforcing damping and no reverse power flow and no high-frequency export at the PCC. Their roles never invert: DRU slope ≈10 mV/A per shelf, aggregated stiffness rising with count; SST slope 5–15× softer.

**Capacitance & Clamps.** Bus capacitance sized $C_{bus} \gtrsim I_{step}\Delta t/\Delta V$ to hold the first dip ≤2 %. Film is distributed at spine, branches, and DRU sidecars. Snubbers or clamps absorb worst-case $L\,di/dt$ from branch-end faults, validated by bench test, not datasheets.

**Reserve & Recharge.** DRUs operate in 50–80 % SoC mid-band, maintaining upward and downward headroom. Floors $R_\uparrow, R_\downarrow$ derive from both power and energy gates across the longest credible burst or MV reconfigure time. Recharge is workload-aware valley-following: admitted only below $P_{avg}$ with MV headroom, ramp ≤50 kW/s per row, hysteresis between L1/L2 to prevent chatter, and instant veto on reserve-floor violation. N+1 redundancy is inviolable.

**MV Compliance.** SST setpoints filtered with $|dP/dt|$ ≤0.1× row nameplate per second, or tighter per relay/tariff. DRU current slew safely exceeds film-cap latency to ensure 2 % limit without overshoot.

**Verification.** Compliance is shown on waveforms, not vendor claims. Disturbance set includes 10–25 % synchronized steps (10–90 s), burst trains near DRU continuous rating, and mixed events with active recharge modulation. Signals logged at ≥50 kS/s (bus, branch) and ≥1 kS/s (DRU power/SoC, SST setpoints), synchronized by PTP. Acceptance requires monotone settling, ≤2 % overshoot, ≥45° equivalent phase margin, and no oscillations in 1–30 Hz band.

Defaults may tighten with silicon and interconnects, but the shape of the contract does not: fast physics at the row, soft energy upstream, explicit droop and ramps, explicit SoC windows, and evidence on traces.

## 5. Fast-Reserve Control Plane (FRP)

The Fast-Reserve Control Plane operates above Tier-0 physics to coordinate reserves, valley-following recharge, and inter-row assistance. It does not guarantee stability; continuity is already preserved by DRU and SST analog control, but it defines how reserves are measured, recharged, and shared when communication is intact.

**Tier-0 Physics.** DRUs operate as current-controlled voltage sources with kilohertz-class loops and stiff droop, arresting fast bus deviations directly. SSTs operate as voltage-controlled current sources with filtered setpoints and softer droop, damping the DC bus and enforcing no reverse power flow and no high-frequency export at the PCC. This hierarchy holds without communication.

**Reserve Floors.** Each row maintains upward and downward reserves $R_\uparrow, R_\downarrow$ above minimum thresholds derived from DRU power, energy, and SoC gates. Floors are never violated; if risk arises, recharge is suspended immediately.

**Recharge Logic.** Recharge is strictly valley-following: admitted only when $P_{row} < P_{avg}$, MV headroom $H_{mv}$ exists, and reserve floors are intact. The command is arithmetic:

$$P_{chg} = \min(H_{mv}, P_{avg} - P_{load} - R_{safety})$$

rate-bounded at $|dP_{chg}/dt| \leq 50$ kW/s per row. Entry requires SoC < $L_1$; exit occurs at $L_2$ or on veto. N+1 redundancy is inviolable.

**Time Discipline.** Devices share a common clock (PTP on TSN or disciplined CAN FD). Reserve commits must act within ≤1 ms one-way; telemetry tolerates 50–200 ms latency. Protection remains outside the network, operating in <100 μs.

**Radialized Independence.** Rows are independent electrical cells at the fast-reserve plane. Because workloads are synchronous means there are no donor / receiver energy transfers; redundancy arises at the MV layer through a radialized distributed topology and FLISR, not through inter-row transfers.

**Failure Handling.** Communication loss or clock drift triggers automatic fallback to Tier-0: fixed droops, conservative reserve floors, no recharge scheduling. Continuity persists, and FRP re-enters softly when restored.

**Verification.** Conformance is defined by waveform evidence. During synchronized bursts, rails must remain inside ±1 % and reserve floors intact. During MV feeder trips, FLISR restores within ~1 s and rows rebalance within a few hundred ms. Under communication loss, the row remains stable in Tier-0 with only slower SoC recovery. Until such traces exist, FRP remains positional.

## 6. Valley-Following Recharge Scheduling

Recharge restores seconds-scale headroom without exposing the medium-voltage system to burst dynamics. It is strictly valley-following: recharge is admitted only when instantaneous row demand lies below the flat average $P_{\text{avg}}$, reserve floors remain intact, and MV headroom exceeds a fixed safety margin. N+1 redundancy is never harvested.

**State Machine.** DRU SoC is held between hysteretic thresholds $L_1 \approx 0.55, L_2 \approx 0.80$. Recharge is admitted when

$$P_{chg} = max\{0, \, min[\, H_{mv}, P_{avg} - P_{load} - R_{safety}\, ]\} \text{ and } |dP_{\text{chg}}/dt| \leq 50 \text{ kW/s per row.}$$

with $|dP_{\text{chg}}/dt| \leq 50$ kW/s per row or tighter by tariff/relay. DRUs sink the recharge command, SSTs track filtered setpoints, racks remain unaffected. Exit occurs at $L_2$ or on veto: reserve-floor violation, MV headroom loss, FLISR topology change, or local fault. Bursts landing mid-valley automatically displace recharge.

**Hierarchy.** DRUs remain the fast bus formers, SSTs enforce average balance under ramp limits, and recharge is always background and conditional. SST bidirectionality may absorb brief rebounds internally, with no reverse power flow and no high-frequency export at the PCC.

**Thermal and Lifetime Margins.** Average DRU discharge must remain within $N \cdot P_{\text{cont}}$. Recharge duty is equivalently budgeted so thermal and coolant setpoints are preserved. SST thermal headroom reduces $P_{\text{chg}}$ directly, ensuring the ±400V rails remain steady under heat stress.

**Acceptance.** Compliance is demonstrated by waveform traces: rails must remain within ±1% during recharge, MV draw must stay within ramp bounds, reserve floors must hold and exit at $L_2$ must be clean. Bursts landing mid-valley must displace recharge with no measurable DC motion. During FLISR events, recharge must freeze immediately, DRUs must bridge, and reconfigured flows must respect segment caps with zero PCC export. Until these traces exist—volts, amps, SoC, and setpoints time-aligned—recharge remains positional.

## 7. MV Distribution, Protection, and FLISR Topologies

The row envelope (tight bus limits, bounded |ΔP| and dP/dt, and no high-frequency export) nests under IEC 61000-3-3 flicker indices and IEEE 519 harmonic limits; it is a pre-conditioner for utility PQ studies, not a substitute.

### 7.1 Protection Doctrine

Protection exists to preserve continuity while allowing the smallest possible piece to fail cleanly. The sequence is deterministic: constrain the first-millisecond energy, interrupt branch faults in microseconds, sectionalize the row in milliseconds only when necessary, and let medium-voltage reconfiguration occur on its own timescale without disturbing the ±400$V_{\text{dc}}$ rails.

**Source Behavior.** DRUs as Current-Controlled Voltage Source (CCVS) and SSTs as Voltage-Controlled Current Sources (VCCS) both current-limit inherently through loop control, bounding fault contribution. Protection then focuses on clamping and interrupting distributed capacitance before sources re-establish regulation.

**Branch Protection.** Each branch incorporates a solid-state, bidirectional interrupter located at the tap, commutating surge energy into a qualified clamp. Trips occur in tens of microseconds by desaturation or slope detection, followed by a mechanical air-gap for cold withstand and lockout/tagout. A semiconductor fuse sits in series as backstop. A branch short therefore trips locally: the silicon interrupts, the clamp absorbs, the air-gap isolates, and the spine deviation remains within limit.

**Row Segmentation.** A hybrid sectionalizer clears in 1–3ms into specified spine inductance. It is slower than branch protection and only arms on coherent multi-branch or bus faults. In ordinary branch faults it does not act, ensuring selectivity.

**Grounding and Arcs.** The bus floats with continuous insulation monitoring. Falling below alarm or trip thresholds triggers alarms or isolation of the smallest island, with supervised reclose. Branch boxes incorporate arc-signature detection cross-checked with $di/dt$, tripping before arcs can persist. Acceptance requires standardized spark tests showing arcs extinguish without reignition.

**Pre-charge and Selectivity.** Pre-charge is universal and validated against golden ΔV profiles; no device closes into an uncharged bank. Time grading is explicit: branch <100μs, segment 1–3ms, MV seconds. Always local silicon first, hybrid second, MV last. Lifecycle margins are embedded in trip thresholds, so coordination persists despite tolerance spreads, sensor drift, and device aging.

**Acceptance.** Branch-end bolted faults must show silicon opening <100μs, clamp voltage within limit, spine deviation ≤2%, and air-gap isolation. Segment faults must split the row within milliseconds while the healthy island regulates. Ground-fault drills must force smallest-island behavior with supervised reclose. Standardized arc tests must extinguish without reignition. In all cases, DRUs must hold the bus inside the 2% / 50ms contract and SSTs must remain within ramp limits. Bench traces, not datasheets, prove conformance.

### 7.2 MV Distribution & FLISR

The medium-voltage layer is deliberately slow and selective. Its function is to carry average energy and reconfigure around faults while the rows themselves enforce fast physics. The topology is radialized but distributed redundant: each block is normally fed radially yet tied into a loop that can reconfigure automatically under FLISR. This provides maximum uptime under fault conditions without requiring rows to exchange energy. Figure 2 illustrates this topology, showing how dual-source medium-voltage feeds integrate with row-level ±400V$_{dc}$ distribution through SSTs and IED-controlled FLISR segmentation.

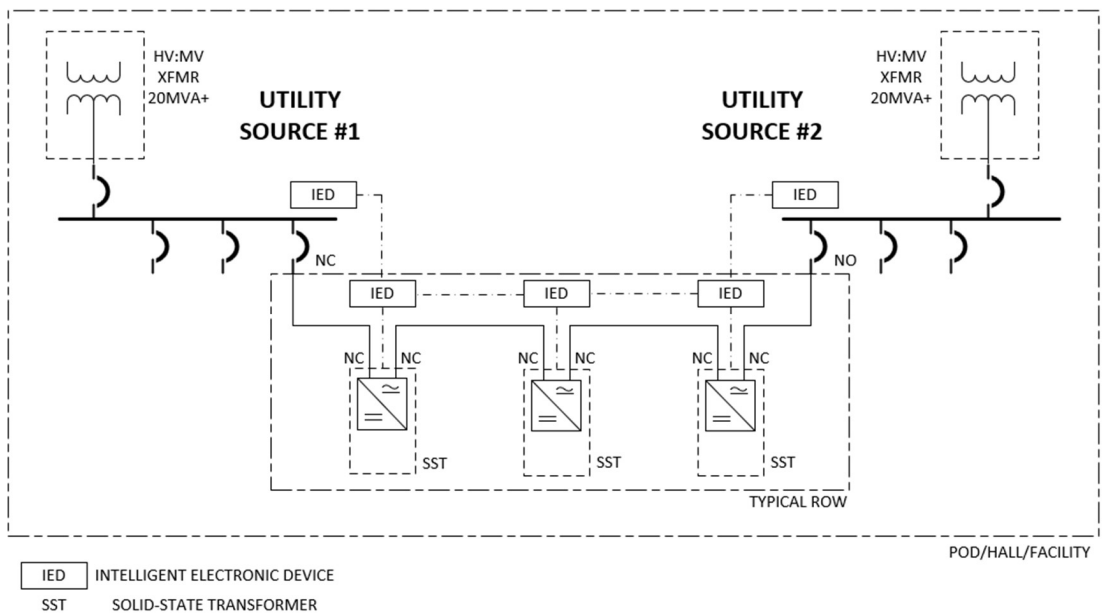

*Figure 2 - Row-scale ±400 Vdc distribution within dual-source MV FLISR.*

**Voltage and Scale.** Operating voltage follows block size and feeder length. At ≤15MVA blocks, 25kV is preferred: a 15MVA block runs ~346A, within thermal comfort bands for conventional metal-clad gear. At ≥18–20MVA or with long pulls, 35/38kV becomes more efficient: ~247A at the same power, reduced current but requiring tighter clearances and often GIS.

**Radialized Redundancy.** The preferred geometry is a loop with FLISR, normally radialized by open ties. A feeder fault trips, and FLISR sectionalizes the fault and recloses the opposite tie in ~1s. Rows ride through on their DRUs as if enduring a training burst. After restoration, the loop remains radialized; live rings are never closed on DC loads. This architecture ensures that each block has two independent MV feed paths yet operates as a stable radial system during normal operation. Above this boundary, traditional availability metrics remain applicable at the facility-distribution layer, where multi-source utility and generator backup govern minute-to-hour resilience.

**Row Independence.** Because AI training loads are synchronous, rows surge together. Each row is responsible for its own fast reserve, and no reverse power flow and no high-frequency export at the PCC. Recharge is suspended during MV events and resumes only after reconfiguration is complete.

**Relaying and Interlocks.** MV protection is written for DC-coupled loads, not rotating machines. Directional overcurrent and reverse-power functions tolerate local dynamics but enforce no reverse power flow and no high-frequency export at the PCC. A tripped feeder remains open until isolation is verified; ties close under supervised logic. SSTs remain blocked until the new topology is confirmed, then ramp under control.

**Scalability and Acceptance.** Adding blocks scales linearly: more feeders at 25kV, or transition to 35/38kV for larger aggregates. In every case, the row contract does not change: DRUs enforce millisecond physics, SSTs provide slow balance, MV carries average energy. Acceptance is proven by feeder trip and FLISR restore traces showing capped segment currents, ≤2% DC deviation, and no reverse power flow and no high-frequency export at the PCC.

## 8. Scaling Contract (Row → Pod → Hall → Campus)

As shown in Figure 2, the MV layer enforces Availability through source redundancy and FLISR reconfiguration, while rows enforce Computational Continuity locally. Scaling builds upward from this invariant boundary.

Scaling is multiplication of an invariant, not reinvention. The invariant is the row: a ±400$V_{dc}$ spine that satisfies the stability contract with DRUs as fast bus formers and SSTs as damped energy sources. Nothing upstream alters that physics. As rows accumulate, higher layers emerge only to pool average power, schedule recharge, and enforce medium-voltage limits. The row's droop, loops, and acceptance criteria remain unchanged.

**Pod Scale.** Two to five rows share a medium-voltage lineup and a supervisory allocator operating on seconds-to-minutes cycles. Each row is summarized by a compact vector of delivered power, fast-reserve headroom, energy in kWh, SoC band, and thermal margin. The allocator computes constant-power targets for SSTs, staggers recharge windows to hold feeders inside ramp limits, and arbitrates setpoints so segment currents remain under their amp caps. Reserve floors per row remain intact, no reverse power flow and no high-frequency export at the PCC; import is subject to a bounded dP/dt envelope, and recharge is always suppressed before disturbing the rails. DRUs carry millisecond physics locally; the scheduler never enters that loop.

**Oversubscription.** Capacity is bounded by energy, not optimism. With pod source rating $P_{MV}$, row nameplate $P_{row}$, $u$ the average training utilization, $r$ the recharge fraction admitted at the feeder, and $l$ the fractional losses, the pod satisfies

$$P_{MV} \geq N(uP_{row} + rP_{row}) + lNuP_{row}.$$

The ratio $P_{MV}/(NP_{row})$ is set by $u + r$ with a small loss term. Realistic $u = 0.7 - 0.85$ keep the safe ratio near unity. Diversity provides margin when workloads decorrelate; with synchronous orchestration, design to the enforced average, not the hoped-for one.

**Hall Scale.** Multiple pods share a common lineup in a looped, FLISR-segmented topology. The hall controller is not a faster brain but a broader allocator. It tracks per-pod vectors and enforces simple rules: segment currents never exceed amp caps, recharge windows never overlap to breach limits, and reserve floors remain intact except under bounded exception. When a feeder trips, FLISR restores in ~1 s. DRUs carry rows through the gap, and the hall allocator redistributes setpoints. If capacity is scarce, trims are surgical: only the affected block is reduced, and only enough to hold the cap. Rows do not perceive hall-level negotiation.

**Campus Scale.** Scaling to tens of pods introduces transmission-level choices but no new small-signal dynamics. With ~15MVA blocks, 25kV remains efficient; at ≥18–20MVA or long pulls, 35/38kV holds currents near 250–330A. A 69kV+ primary can step down to multiple 25 or 35kV feeders; each block remains an invariant electrical cell. Medium-duration storage, if deployed, resides on the MV side as a schedulable generator to flatten tariffs and arbitrate recharge contention, but never enters the millisecond domain. MVDC is justified only for kilometer-scale feeders with standardized DC breakers; otherwise, AC remains the scaling medium. At this campus-distribution boundary, traditional Availability metrics again apply, governed by multi-source utility supplies and generator backup that ensure minute-to-hour resilience.

**Operational Scale.** Scaling multiplies playbooks, not tuning knobs. A golden row is validated on hardware-in-the-loop benches with droop slopes, loop bandwidths, capacitance placement, and acceptance traces locked. That row becomes the invariant unit. A golden pod validates FLISR ride-through and recharge staggering; it too becomes a unit of replication. Spares scale linearly: branch interrupters, one sectionalizer per spine segment, ~10% DRU shelf reserve, and one SST stage per lineup. Telemetry remains time-aligned everywhere, so post-mortems resolve to physics, not debate.

**Safety Scale.** Grounding, pre-charge, and protection do not change with scale. The ±400V bus remains floating with monitored insulation, supervised recloses, and no device closing onto an uncharged bank. Protection is certified at the assembly level, not piecemeal. MV relays remain conventional except for tolerating limited internal reverse power while no reverse power flow and no high-frequency export at the PCC; import is subject to a bounded dP/dt envelope.

**The Law.** Seconds belong to the row, minutes and hours to MV, averages to schedulers and storage. Physics freezes at the row, math stays explicit at the pod, amp caps are enforced at the hall, and the utility interface remains quiet at the campus. Under this discipline, scaling is replication, not discovery. Thirty rows behave like three, and ninety megawatts is only the sum of thirty tested units. Until golden-unit traces exist, this law is positional; once they do, scaling is inevitable.

## 9. Conclusion: A Contract for Continuity

The challenge raised by *Choukse et al.* is not simply that AI training surges destabilize constant-power loads, but that existing infrastructure lacks a physics-anchored framework to contain them. The response defined here is not an optimization or a tuning exercise; it is a contract. The row is the invariant. DRUs stabilize the bus in microseconds, SSTs regulate and supply energy in seconds, film capacitance and clamps absorb the first edge, and reserve floors guarantee bidirectional headroom. These are measurable obligations, not abstractions: ±1% steady band, ≤2% transient, ≤50 ms return, ≥45° margin, reserve floors intact.

Scaling preserves this invariant. A pod adds only a scheduler that allocates averages and staggers recharge. A hall adds only FLISR and simple amp caps across feeders. A campus adds only transmission-level arithmetic. At no stage is the row contract diluted, nor are its acceptance tests retuned. Thirty rows therefore behave like three, and ninety megawatts remains only the sum of thirty identical and proven units.

The doctrine is that protection and stability precede software. Physics works first; control planes merely allocate energy later. The acceptance of this architecture is not brochures or simulations but waveforms: traces of branch faults clearing in microseconds, rows holding inside the 2%/50 ms window, MV loops restoring under FLISR with no reverse power flow and no high-frequency export at the PCC; import is subject to a bounded dP/dt envelope, and synchronized bursts absorbed without ripple. If those waveforms exist, the system is conformant; until then, it remains a positional architecture.

What this paper asserts, therefore, is that the way forward is not to reinvent stability for every scale, but to freeze it at the row and replicate. By codifying invariants, bounding dynamics, and demanding empirical proof as the technology matures, the architecture transforms the AI power problem from an existential uncertainty into an engineering specification. Computational Continuity is preserved not by hope, but by contract.